\documentclass[11pt]{article}

\usepackage[a4paper,margin=1in]{geometry}
\usepackage{microtype}

\usepackage[utf8]{inputenc}
\usepackage[T1]{fontenc}

\usepackage{amsmath}
\usepackage{amssymb}
\usepackage{mathtools}

\usepackage{graphicx}
\usepackage{subcaption}
\usepackage{booktabs}
\usepackage{threeparttable}
\usepackage{tablefootnote}
\usepackage{makecell}
\usepackage{array}
\usepackage{adjustbox}
\usepackage{stfloats}

\usepackage{placeins}
\usepackage{cuted}
\usepackage{capt-of}

\usepackage{fvextra}
\usepackage{listings}
\usepackage{mdframed}

\usepackage{xcolor}
\usepackage{enumitem}

\usepackage{algorithm}
\usepackage[noend]{algpseudocode}

\usepackage{tikz}
\usepackage{pgfplots}
\pgfplotsset{compat=1.18}

\usepackage{textcomp}
\usepackage{lipsum}
\usepackage[toc,page]{appendix}

\usepackage{authblk}

\usepackage[
  colorlinks=true,
  linkcolor=blue,
  citecolor=blue,
  urlcolor=blue
]{hyperref}

\title{
\textcolor[HTML]{a07bcc}{MOONWALK}:
\textcolor[HTML]{a07bcc}{M}ediating
\textcolor[HTML]{a07bcc}{O}perati\textcolor[HTML]{a07bcc}{on}s
\textcolor[HTML]{a07bcc}{w}ith Intent--Evidence--Action
\textcolor[HTML]{a07bcc}{Al}ignment Across Junior--Supervisor Review
Wor\textcolor[HTML]{a07bcc}{k}flows in Animation/VFX Pre-Production
}

\author[1,2]{Shih-Yu Lai}
\author[3]{Wen-Fan Wang}
\author[2]{Sai Ling}
\author[2]{\\Shaune Jan}
\author[1]{Bing-Yu Chen}
\author[4]{Xiang `Anthony' Chen}

\affil[1]{National Taiwan University, Taipei, Taiwan}
\affil[2]{MoonShine Animation Studio, Taipei, Taiwan}
\affil[3]{Cornell Tech, New York City, USA}
\affil[4]{HCI Research, UCLA, Los Angeles, California, USA}

\affil[ ]{\small
\texttt{akinesia112@gmail.com},
\texttt{vann@cmlab.csie.ntu.edu.tw},
\texttt{saix34@gmail.com},
\texttt{shaune.jan@gmail.com},
\texttt{robin@ntu.edu.tw},
\texttt{xac@ucla.edu}
}

\begin{document}

\maketitle

% ── Abstract ─────────────────────────────────────────────────────
\begin{abstract}
Animation and VFX pre-production review requires teams to translate loosely specified
creative intent---briefs, evolving specifications, heterogeneous references, and verbal
decisions---into revisions that junior artists can execute without repeated clarification.
In practice, criteria drift across iterations, review judgments lose their evidential basis,
and the reasoning behind a request rarely survives the senior--junior handoff. We
contribute a \textbf{design framework for intent--evidence--action alignment}: intent is articulated into a shared project record, judgments are anchored to grounded evidence, and authorized decisions are converted into clear revision tasks tied directly to reference notes. We instantiate this framework in \textbf{MOONWALK}, a professional
pre-production review system comprising a shared intent record, reference/specification
anchoring, structured work-in-progress comparison, and supervisor-authorized action
planning. In this workflow, AI handles administrative coordination—flagging missing context and organizing notes—while artists retain full creative direction. An \textbf{in-studio study} with professional practitioners compares MOONWALK with
a chat-only (chatbot) interface using matched production materials, while
participants' existing workflows provide a retrospective ecological
baseline. Results indicate stronger intent alignment, decision
traceability, and checklist executability, while also showing that aesthetic authority and
final prioritization must remain with practitioners. The evaluation establishes the value
of the integrated structured workflow over unstructured conversational AI chatbot.
\href{https://github.com/Akinesia112/Moonwalk/tree/english-version}{{\textbf{Code}: \textbf{https://github.com/Akinesia112/Moonwalk/tree/english-version}}}
\end{abstract}

% ── Keywords ─────────────────────────────────────────────────────
\begin{center}
\end{center}

\vspace{0.5em}

% ── Hero / Teaser Figure ─────────────────────────────────────────
\begin{figure}[ht]
  \centering
  \includegraphics[width=\textwidth]{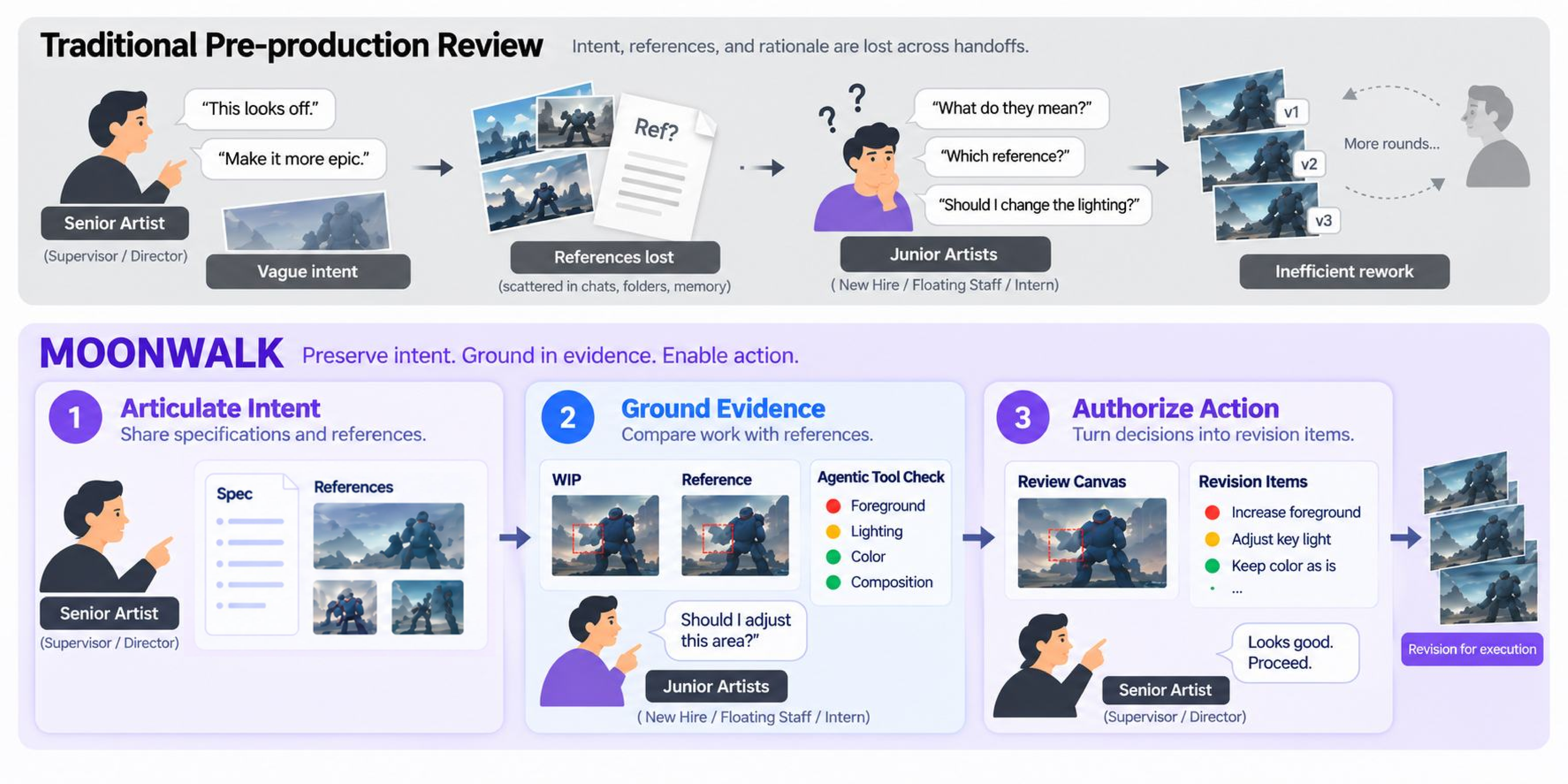}
  \caption{
  Pre-production review in 2D animation and VFX often breaks the
  \textit{intent--evidence--action chain}, as creative intent, supporting
  references, and revision rationale are lost across senior--junior handoffs
  in their Work in Process (WIP) artwork.
  MOONWALK maintains this chain by preserving intent, grounding review in
  inspectable evidence, and converting supervisor-authorized decisions into
  evidence-linked, executor-ready actions for revision items.
  The framework bounds AI to coordination---requesting missing evidence,
  reminding either role of stated requirements, and organizing authorized
  items---not to aesthetic judgment.
  }
  \label{fig:teaser}
\end{figure}

% NOTE:
% ACM's \Description{} has no standard arXiv/article equivalent.
% Keep the alt-text description as a source comment if desired:
%
% The left side depicts a traditional pre-production review workflow in which
% a senior artist and a junior artist repeatedly exchange vague comments and
% clarification questions, producing an iterative rework loop. The right side
% depicts MOONWALK as an instantiation of an intent--evidence--action chain.
% The Intent component contains a shared specification and annotated visual
% references. The Evidence component connects artwork observations and the
% junior artist's interpretation to specific references, specification clauses,
% and prior decisions, allowing both roles to inspect the same rationale.
% The Action component contains a prioritized revision plan in which each
% requested change cites its evidential basis and is authorized by the
% supervisor. AI assistance appears as a supporting layer for prompting,
% reference comparison, and information organization rather than as the source
% of final aesthetic judgment.

% ── Main text ────────────────────────────────────────────────────
\section{INTRODUCTION}

Pre-production review in 2D animation and VFX is a repeated coordination process: a
\textbf{director} provides a brief, references, or visual goals; a \textbf{supervisor}
interprets that direction while reviewing successive versions; and a \textbf{junior
artist} revises the work. Existing production tools already organize shots, assets,
versions, notes, and approvals, and recent AI systems can generate comments or compare
visual material. Yet the difficult part of studio review is often neither storing feedback
nor producing more of it. It is preserving what the supervisor meant, what evidence
defines that intent, and how the junior artist should act on it across iterative handoffs
\cite{griffithCHI24,feedbackByDesign2026}.

Consider the review in Figure~\ref{fig:teaser}. A senior artist says the scene looks off'' and asks a junior to make it more epic,'' without specifying which reference or visual property matters. The junior may revise the lighting, only to learn that the supervisor meant something else. In another case, a director said ``I want that feeling'' while showing a landscape image with data-visualization overlays; the intended feature was the cyberpunk annotation language, leading the artist to search for the wrong references for days. In both cases, the intent was conveyed, but the evidence needed to interpret it was not.

\begin{figure}[t]
\centering
\includegraphics[width=\linewidth]{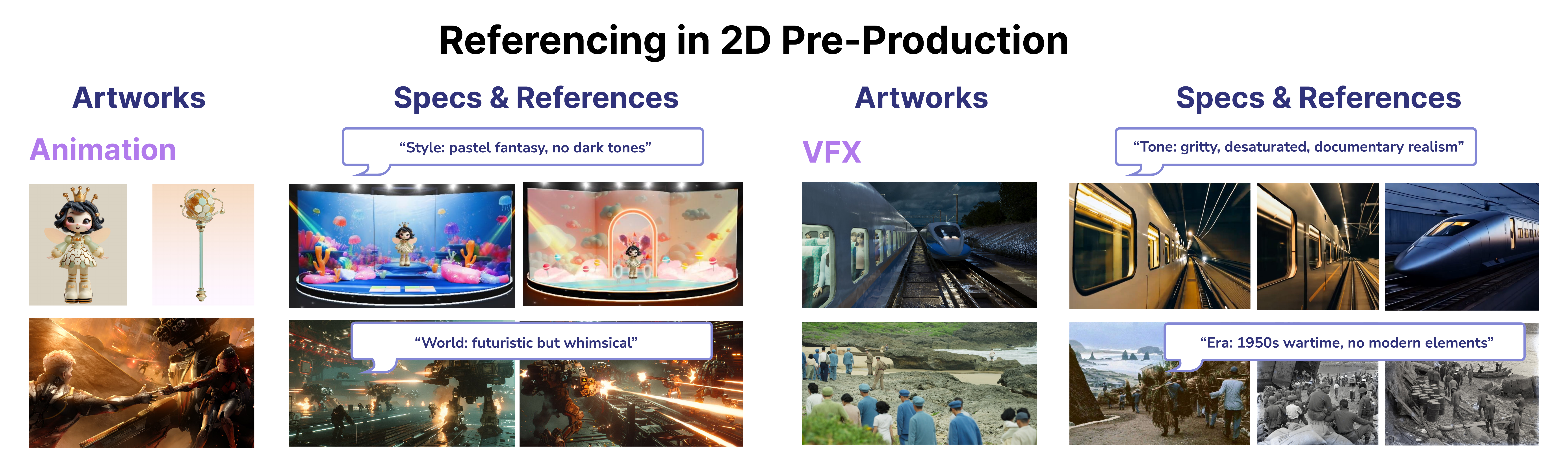}
\caption{Referencing in 2D pre-production. Artworks establish visual targets, while
specifications and references bound the acceptable exemplar space. Current practice
rarely preserves how a particular reference region or specification motivates a
revision request.}
\label{fig:referencing}
\vspace{-0.5cm}
\end{figure}

Figure~\ref{fig:referencing} illustrates two artifact types that already support review:
\textit{artworks} that establish visual targets and \textit{specifications and references}
that constrain what those targets mean. These artifacts are common in practice, but
their relationships are rarely preserved. A reference may be stored without recording
which region matters; a specification may disappear from later review; and a revision
note may survive without the rationale that motivated it. Small interpretive differences
therefore accumulate into criteria drift and repeated clarification
\cite{areWeOnTrack2025,designingTemporalWork2025}.

We describe this breakdown as a fragile \textbf{intent--evidence--action chain}
(Figure~\ref{fig:teaser}). \textit{Intent} is the active creative goal and constraints;
\textit{evidence} is the reference, specification, prior decision, or observable property
that supports a review judgment; and \textit{action} states what should change and what
would count as resolution. When these links are lost, feedback becomes a verdict such as
``it feels off'' rather than a grounded request that another role can execute.

Repairing this chain requires better supervisor--artist communication rather than a more autonomous reviewer. Supervisors must make missing specifications, references, and the relevant object, region, or property explicit; junior artists must be able to inspect those anchors, resolve ambiguity, and record intentional deviations. That interpretation then travels with the work into the next review. AI supports this exchange through evidence-bounded checks and clarification prompts, abstaining when the project record does not specify a criterion.

Prior HCI work addresses related breakdowns at different points in collaborative work. Scott et al. support prospective reflection by helping meeting participants articulate purposes, anticipated challenges, and success conditions before a meeting \cite{whatDoesSuccessLook2025}. Vanukuru et al. extend this across recurring meetings by reconnecting retrospective and prospective work over time \cite{designingTemporalWork2025}. Sharma et al. examine feedback between users and conversational agents, identifying failures of common ground, verifiability, communication, and informativeness \cite{feedbackByDesign2026}. MOONWALK instead targets an asymmetric production handoff: a senior practitioner’s judgment must remain linked to the reference, specification, or prior decision that supports it, and then be translated into revision work that a junior practitioner can execute.

We contribute an \textbf{intent--evidence--action framework} for review handoffs. It links \emph{intent} (active goals and constraints), \emph{evidence} (references, specifications, prior decisions, or observable artifact properties supporting a judgment), and \emph{action} (an authorized revision with target, priority, and completion condition). These links persist across iterations, allowing each revision to be traced from action to supporting evidence and active intent. From this abstraction, we derive three design goals: preserve intent across iterations, make review evidence inspectable and traceable, and retain these links when decisions become executable revision work.

\textbf{MOONWALK} instantiates this framework with AI limited to coordination: requesting missing references or specifications, pulling up the corresponding reference image or specification clause, checking stated requirements against submitted work, surfacing unresolved requirements, and organizing supervisor-authorized revisions. AI does not introduce aesthetic criteria or independently judge lighting, color, composition, or style beyond the project record and explicit human judgment.

We evaluate MOONWALK in an in-studio, within-subject study with 19 professional animation and VFX practitioners (10 senior-role, 9 junior-role) across two studio contexts. Participants compared MOONWALK with a Chat-Only interface using the same AI, while existing studio workflows served as a retrospective ecological comparison. MOONWALK was rated above neutral on 12 of 13 Likert items and led eight of nine comparative questions, including junior-executable checklists (74\%), reduced senior--junior clarification (63\%), and blind-spot identification and evidence-linked feedback (79\% each); perceived control was the main exception. Our contributions are:
\begin{itemize}
    \item \textbf{An intent--evidence--action design framework} for preserving creative
    intent, inspectable evidence, and executable revision action across iterative
    senior--junior handoffs.

    \item \textbf{MOONWALK, an working prototype built on these principles} that structures
    supervisor intent, evidence, junior interpretation, and supervisor-authorized action
    while keeping aesthetic judgment with practitioners.

    \item \textbf{Empirical evidence from professional practitioners} through \textsc{in-studio} formative study and an summative evaluation examining
alignment, traceability, actionability, clarification needs, and human agency.
\end{itemize}

\section{RELATED WORK}
\label{sec:RelatedWorks}

% We organize prior work around the
% coordination problem and the structured workflow. AI-based feedback
% is discussed as an implementation method within these lines of work rather than as an
% independent theoretical contribution.

\subsection{Creativity Support in Production Contexts}

Creativity-support systems make alternatives inspectable and revision histories
recoverable across ideation and production. Reference-based systems connect exemplars
to output dimensions: CreativeConnect maps reference elements to design decisions
\cite{creativeconnect2024}; StyleFactory supports ranking-based style control
\cite{stylefactory2024}; and MemoVis anchors asynchronous feedback to visual evidence
\cite{memovis2024}. Systems for large reference spaces similarly show that continuity
matters: AIdeation and GenTune support cross-reference traceability
\cite{aideation2025,GenTune}; POET reduces drift through personalization loops
\cite{poet2025}; and ImaginationVellum preserves prompt--stroke--output histories
\cite{imaginationvellum2025}. These systems motivate persistent reference records, but
they do not address the full senior--junior review handoff in which the rationale behind
a reference must become executable production work.

Animation, motion, and video-authoring research further establishes the value of
inspectable intermediate structures. Timeline-and-component models separate motion
from content \cite{reflectingAnimatedDataVideoCHI25}; keyframe systems
\cite{dataanimatorCHI21}, visualization grammars \cite{dataparticlesCHI23}, layered
3D authoring \cite{layeredStylized3D2022}, document-to-video workflows
\cite{doc2video2022}, and sculpting autocomplete \cite{autocompleteSculpting2020}
make intermediate decisions visible enough to verify and revise. Production-proximate
AI systems extend this principle through visually grounded code and repair
\cite{logomotionCHI25}, versioned agent workflows \cite{mapstory2025}, auditable task
decomposition \cite{editduet2025}, local compositional editing
\cite{compositionalStructures2025}, and structured narrative assets
\cite{collaposer2026}. MOONWALK adopts this focus on transparency, extending it to cross-role handoffs from creative direction to evidence-linked revision.

\subsection{Collaborative Creativity Tools}

Recent collaborative creativity systems increasingly treat creative
intent as an evolving representation rather than a one-time prompt or
brief. Such systems externalize goals, intermediate interpretations, and
decision histories so that collaborators can revisit and revise them
across iterations. Work on temporal support for recurring collaboration
shows that interfaces must connect prior decisions with current activity
rather than merely summarize isolated sessions
\cite{designingTemporalWork2025}. Research on collaborative evaluation
similarly emphasizes making success criteria explicit and maintaining
traceable relationships between goals, evidence, and subsequent
decisions \cite{whatDoesSuccessLook2025,feedbackByDesign2026}.

This requirement becomes more important when generative AI participates
in creative work. AI can help teams organize information and surface
candidate interpretations, but access to generated content does not by
itself establish shared standards. Studies of collaborative prompting
show that human partners continue to rely on human--human discussion and
shared expertise when negotiating creative direction
\cite{whenTeamsEmbraceAI2024}. MOONWALK extends these efforts by
maintaining a persistent relationship among the active intent,
reference- and specification-based evidence, and the revision actions
authorized by practitioners across the senior--junior review handoff.

HCI systems have addressed adjacent collaborative breakdowns through shared criteria,
visible traces, and inspectable disagreement. Prior work surfaces success criteria before
discussion \cite{whatDoesSuccessLook2025}, visualizes otherwise invisible meeting
structure \cite{meetmap2025}, prevents premature consensus through issue mapping
\cite{EchoMind}, preserves continuity across sessions
\cite{designingTemporalWork2025}, and supports productive disagreement
\cite{reflexis2026}. Research on feedback quality further shows that failures often arise
from missing common ground and unverifiable claims rather than from individual
deficiencies \cite{feedbackByDesign2026}. Power asymmetries also matter: technically
actionable information may not change behavior when authority and accountability are
unclear \cite{opportunitiesBarriersAImeeting2026}. These challenges highlight the need for a transparent, single record shared equally between supervisors and junior artists.

AI can assist this coordination, but more assistance is not automatically better.
Shared AI displays and role-specific agents can broaden coverage
\cite{ladica2025,perspectra2025,postermate2025,towardsAIcolleagues2026}, while excessive
AI assistance can reduce cognitive engagement \cite{chen2025assistance}.
Agentic interfaces therefore require steering, inspection, rollback, and explicit human
authorization \cite{agdebuggerCHI25}. MOONWALK adopts these controls but positions the
AI as a production-coordination aid that prompts, retrieves, compares, and drafts; the system centers on maintaining the shared review history rather than automated generation.

\subsection{Intent Alignment in Structured Production Workflows}

Previsualization and production-tracking systems already structure substantial portions
of filmmaking and animation work. CineVision gives directors and cinematographers a
shared, editable previsualization storyboard \cite{CineVision}; related systems support
generative previsualization from rough 3D \cite{PrevizWhiz}, mobile shot planning and
capture \cite{CineCraft}, and storyboard retrieval from visual-intent canvases
\cite{CineMuse}. These systems align creative leads before or during shot planning,
typically by making the planned shot itself the shared artifact. MOONWALK targets a downstream phase in the review process: once direction exists, how does its rationale persist
through iterative review and become work that another role can execute?

Commercial production-tracking platforms such as ShotGrid/Autodesk Flow Production
Tracking already support shots, assets, tasks, versions, notes, annotations, review, and
approval. These platforms already give production work its structure, and MOONWALK
does not replace them. Current tracking platforms log comments, but fail to capture the visual rationale connecting a specific reference to a revision request: a note can be
stored without preserving which specification clause, reference region, or prior
decision justified it; a junior artist's interpretation can remain invisible until it
appears as incorrect work; and a list of comments can lack the priorities and completion
conditions required for independent execution. MOONWALK acts as an annotation layer that augments existing pipeline tracking tools with intent traceability.

Multimodal models provide one way to scale this layer, but the system's scope and limits must be clearly defined. LLM-as-a-judge and MLLM evaluation enable scalable comparison
\cite{surveyLLMJudge2024,mlljudge2024}, while human-in-the-loop systems emphasize
user-defined criteria, agreement inspection, and active auditing
\cite{pan2024evalllm,gebreegziabher2025metricmate,shankar2024validators,
prometheusvision2024}. Evidence indicates that MLLMs are more reliable for
\emph{observational} tasks such as perceptual reference matching than for
\emph{interpretive} judgments requiring tacit domain standards
\cite{smes2025,aligningMLLMexperts2026,visjudgebench2026,artmentor2025,criticv2025}.
Accordingly, MOONWALK uses model output to surface candidate discrepancies and missing
information, while supervisors retain aesthetic interpretation, prioritization, and final
authorization. We contribute a structured workflow that links artistic feedback directly to verifiable, reference-backed revision tasks.

\section{FORMATIVE STUDY}

We conducted a formative study to investigate how creative intent, quality criteria, and
revision evidence are articulated across the artist--supervisor review cycle in animation
and VFX production, and where their relations break down. We recruited 12 practitioners
(1--16 YoE, Mean = 4.96) across two studios: directors ($n=2$), a supervisor
($n=1$), senior and junior artists ($n=8$), and a PM ($n=1$), spanning commercial
advertising, character animation, virtual production, and film/TV VFX. Sessions lasted
30--60 minutes; interviews were audio-recorded, transcribed, and analyzed using thematic
analysis \cite{braun2006thematic}. Full participant details and procedure are provided in Appendix~C; the interview is in Appendix~D.

The studios used in-person review, production-tracking notes, chat, shared documents,
calls, and annotated images. Senior practitioners carried final responsibility for
aesthetic coherence, while junior practitioners often had to interpret terse instructions
without repeatedly interrupting supervisors. Authority structure, communication norms, and tool ecology shape both the observed breakdowns and the transferability of our design implications.

\subsection{Findings}

\subsubsection{Intent Did Not Survive the Handoff as a Shared Interpretation.}

Directors and supervisors frequently reviewed work by walking to a workstation and
pointing at the screen. These exchanges conveyed rich context but left little record
(P6, P8, P9). Written feedback in ShotGrid or Zulip was often too sparse to reconstruct
what had been indicated visually: one participant noted that even a detailed document
omitted the reference region shown during the conversation (P8). Feedback arriving
through disconnected channels also obscured which instruction superseded another
(P3, P11, P15, P16, P18). At the same time, junior artists' interpretations and reasons
for a visual choice were rarely returned to supervisors in a structured form. The
supervisor often discovered that interpretation only after it had materialized as
incorrect work. The handoff was therefore lossy in one direction and absent in the other.

\subsubsection{Review Judgments Were Delivered as Verdicts Rather Than Grounded Evidence.}

Participants across roles described judgments that did not identify the reference,
specification clause, prior decision, or visual property that motivated them. Supervisors
and directors routinely used phrases such as ``feels off,'' ``the atmosphere is wrong,''
or ``not magical enough'' (P2, P6, P8, P9, P12, P13). Junior artists cycled through
plausible interpretations with no narrowing signal (P12, P13, P16--18). Two junior artists described a client calling an image ``ghostly'' without clarifying whether this meant
color grading, shadow depth, or compositional density; each attempted interpretation was
rejected without a more specific rationale (P12--13). The PM similarly received
emotional reactions such as ``this is ugly'' and had to translate them into production
work (P15). The decisive criteria existed, they simply remained in reviewers' mental models, unanchored to any shared artifact (P3, P8, P9, P12, P15).

\subsubsection{Criteria and Revision Obligations Drifted Across Iterations.}

Participants described standards that silently changed or became impossible to
reconstruct because no persistent rationale connected successive versions. A CG lead
characterized this as a recurring onboarding problem: junior artists could not see the
gap between their work and the reference (P3). The CG lead and supervisor converged on
two failure modes: an \textit{absent reference}, in which the artist did not know the
target, and an \textit{undetected gap}, in which the artist believed the target had been
matched (P3, P9). The PM further noted that, without a record of agreed direction,
legitimate client revisions became indistinguishable from contradictions (P15).
Participants also received undifferentiated lists of comments that did not distinguish
blocking changes from optional refinements, prompting further clarification before work
could begin.

\subsection{Design Goals}

Our study revealed a core workflow breakdown: studios lack a persistent, shared
representation that preserves the relation among creative intent, review evidence, and
executable action across both directions of the senior--junior handoff. We derive three design goals aligned with the chain in Figure~\ref{fig:teaser}.

\begin{itemize}
    \item \textbf{DG1---Persistent Intent Articulation:}
    Because verbal intent and junior interpretations did not survive handoff, the
    workflow should preserve project goals, constraints, uncertainty, and both roles'
    interpretations in a shared record that remains available across versions.

    \item \textbf{DG2---Evidence Anchoring and Traceability:}
    Because judgments arrived as verdicts, every candidate issue and authorized
    revision should identify its evidential basis---a reference region, specification
    clause, prior decision, or observable artifact property.

    \item \textbf{DG3 – Actionable, Reference-Grounded Tasks:}
    Because criteria and obligations drifted, each final revision item should state
    what to change, why the change follows from the evidence, its priority, and what
    would count as resolution.
\end{itemize}

Both roles require full visibility: supervisors must see how junior artists interpret feedback, and artists must see the reasoning behind senior requests. Supervisors must be able to inspect how junior artists interpreted
the brief and references; junior artists must be able to inspect the rationale behind
senior requests; and either role must be able to flag uncertainty or request additional
evidence before action is finalized.

We operationalize these design goals in MOONWALK through unified specifications, annotated reference hubs, and structured feedback checklists. Persistent specifications,
annotated references, interpretation records, and evidence-linked action items are the core features. AI can support these mechanisms, but the goals neither presuppose AI nor
grant it aesthetic authority.
Section~4 describes how MOONWALK instantiates these goals.
\section{SYSTEM DESIGN \& IMPLEMENTATION}
\label{sec:system}

\begin{figure*}[t]
  \centering
  \includegraphics[width=\textwidth]{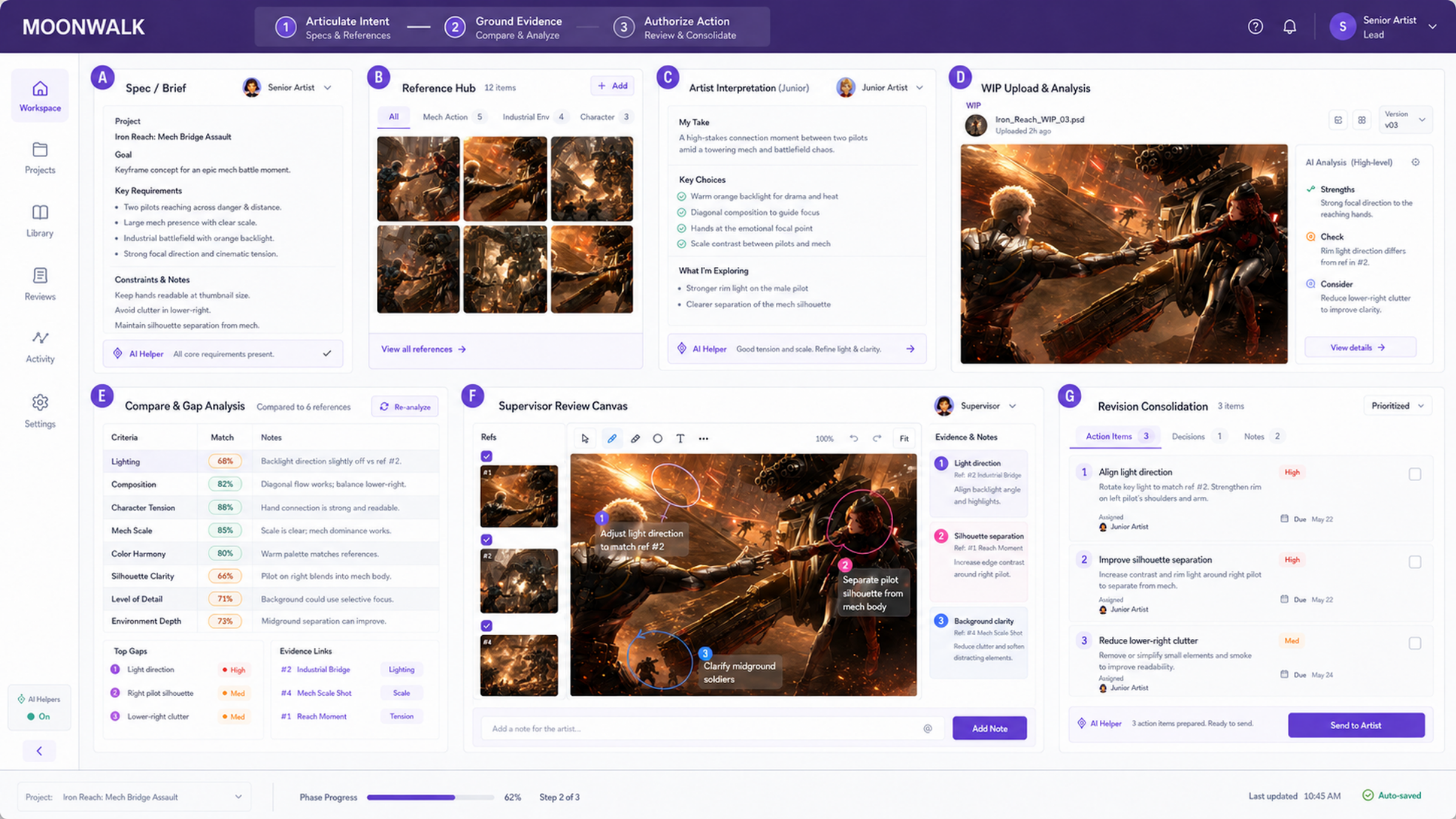}
  % \caption{\textsc{MoonWalk} interface panels across the IR4IA workflow.
  % \textit{\textcolor[HTML]{a07bcc}{Intention Reflection}}:
  % (a)~Spec/Brief structures creative intent into seven fields;
  % (b)~Reference Hub organizes exemplars by category and priority with per-reference intent notes.
  % \textit{\textcolor[HTML]{a07bcc}{Artwork Reflection}}:
  % (c)~Artist Self-Reflection externalizes working intentions before upload;
  % (d)~Upload \& Analysis streams eleven-dimension agent debate results in real time;
  % (e)~Compare \& Gap Analysis surfaces discrepancies between artworks and references ---\textcolor[HTML]{EA4040}{red} for first-order corrections, \textcolor[HTML]{EAA404}{yellow} for second, and \textcolor[HTML]{359D8A}{green} for pass.
  % \textit{\textcolor[HTML]{a07bcc}{Feedback Reflection}}:
  % (f)~Supervisor Review Canvas supports pixel-level annotation with version history;
  % (g)~Conflicts Resolve produces a U1/U2/U3 prioritized revision checklist from supervisor, client, and AI input;
  % (h)~Intent Alignment Assistant provides five role-differentiated AI agents throughout the workflow.}
\caption{
MOONWALK interface panels across three review operations.
\textbf{Articulate Intent:}
(a) \textit{Spec/Brief} records the active project goals, requirements, and
constraints;
(b) \textit{Reference Hub} organizes visual exemplars with per-reference
intent notes; and
(c) \textit{Artist Interpretation} captures the junior artist's reading of
the brief, references, and intentional deviations before review.
\textbf{Ground Evidence:}
(d) \textit{WIP Upload \& Analysis} evaluates the submitted work through
eleven analysis dimensions against available project evidence; and
(e) \textit{Compare \& Gap Analysis} presents candidate discrepancies
together with their supporting specifications or references.
\textbf{Authorize Action:}
(f) \textit{Supervisor Review Canvas} supports region-level annotation and
review; and
(g) \textit{Revision Consolidation} organizes supervisor-authorized
decisions into a prioritized, evidence-linked checklist for junior artists.
}
\label{fig:ui}
\end{figure*}

MOONWALK carries one project record across pre-production review: the active brief,
annotated references, the junior artist's stated interpretation, the submitted work,
supervisor annotations, and authorized revision items. The interface follows this record
through three operations---\textit{Articulate Intent}, \textit{Ground Evidence}, and
\textit{Authorize Action}---so that information entered before review remains visible when
feedback is produced. AI assists communication within these operations by requesting
missing context, comparing a WIP with supplied anchors, and organizing review material.
The framework treats supplied specifications, references, prior decisions, and explicit
human judgments as the source of review criteria.

\subsection{Workflow and Interface}

Figure~\ref{fig:ui} summarizes the three operations over the same project record. The
first makes the intended target explicit, the second places the current work beside the
evidence used to assess it, and the third converts supervisor decisions into work that a
junior artist can execute. The Spec/Brief, Reference Hub, comparison workspace, Review
Canvas, and checklist are therefore connected views of one review history rather than independent tools.

\subsubsection{Articulate Intent}

Review begins in the \textbf{Spec/Brief} and \textbf{Reference Hub}. Supervisors record
the active direction and annotate what a reference is intended to communicate: the
relevant object or region, the property to follow, and any content that should be ignored.
The junior artist reads the same record before submission and can note which requirements
were followed, answer or flag unresolved questions, and explain intentional deviations in
the \textbf{Artist Interpretation} panel. These notes are stored with the WIP so the next
supervisor review includes both the artifact and the artist's account of how the current
direction was understood.

\paragraph{Example in practice.}
If a supervisor writes ``make the mech feel more battle-damaged'' without identifying a
usable target, AI can ask for a reference or specification and prompt the supervisor to
mark the relevant object or region. Once the supervisor states ``use the beam-scorch marks
in Reference~\#2; ignore the background,'' that clarification becomes available to the
junior artist and to later review steps.

\subsubsection{Ground Evidence}

The junior artist uploads the WIP to a comparison workspace that also shows the active
specification, relevant reference notes, prior decisions, and the artist's interpretation.
The \textbf{Compare \& Gap Analysis} surfaces candidate differences, while the
\textbf{Review Canvas} lets the supervisor mark the region that motivates a comment. For
the framework, a useful automated observation is one that can be checked against an
explicit project anchor---for example, whether a required object is present, whether a
specified relation holds, or whether an observable property differs from an annotated
reference. The observation is presented with its supporting context for supervisor
inspection; it does not itself authorize a revision.

\paragraph{Example in practice.}
Suppose the brief requires a visible antenna array and Reference~\#3 marks its placement,
but the WIP omits it. AI can surface the omission and cite both anchors. The junior artist
can respond that the antenna was intentionally removed because of a newer client note;
that explanation remains attached to the submission for the supervisor to resolve.

\subsubsection{Authorize Action}

The supervisor decides which observations should become production work. In the
\textbf{Review Canvas} and \textbf{Revision Consolidation} view, accepted items retain
their supporting reference, specification, prior decision, supervisor annotation, or
client instruction. The supervisor can rewrite or merge items, assign urgency, and state
a completion condition. The resulting checklist is the handoff artifact returned to the
junior artist.

\paragraph{Example in practice.}
After the supervisor marks the missing antenna as blocking and annotates its intended
placement, AI can consolidate duplicate comments into ``add the antenna array at the
location marked in Reference~\#3; preserve the current body silhouette.'' If an item has
no stated rationale, the interface can prompt the reviewer to supply one before the item
is finalized.

\begin{figure*}[htbp]
    \centering
    \includegraphics[width=0.98\linewidth]{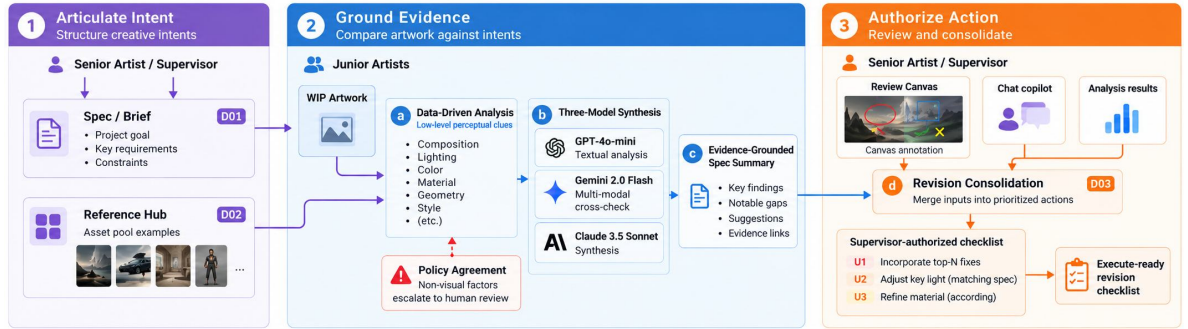}
    \caption{MOONWALK system architecture across the three operations.
\textbf{Articulate Intent:} the supervisor structures a Spec/Brief (DG1)
and annotated Reference Hub (DG2). \textbf{Ground Evidence:} (a) eleven
dimension agents score the WIP (eleven in total; six shown); non-visual factors, high-disagreement, low-score dimensions escalate to human review;
remaining results feed (b) a three-model synthesis (GPT-4o-mini: visual
observation; Gemini 2.0 Flash: specification and reference comparison;
Claude 3.5 Sonnet: synthesis) yielding (c) a reference-grounded spec
summary. \textbf{Authorize Action:} canvas annotations, client opinions,
and analysis results merge in the (d) Revision Consolidation view,
outputting a U1/U2/U3-prioritized, supervisor-authorized checklist (DG3).}
    \label{fig:pipeline}
    \vspace{-0.5cm}
\end{figure*}

\subsection{Technical Implementation}
\label{sec:technical-implementation}

The evaluated prototype organizes automated analysis into eleven dimensions
(Figure~\ref{fig:pipeline}). Seven are WIP- and evidence-facing:
lighting, composition, color, style, perceptual quality, sketch/line
quality, and specification faithfulness. Their judge functions combine
pixel-derived signals such as sharpness and artifact statistics, saliency
and layout, white-balance/exposure cues, and CLIP-derived style or prompt
alignment with lexical overlap and relevance against the supplied brief,
reference, and reflection text. The remaining four---controllability,
consistency, efficiency, and stability---are auxiliary implementation
diagnostics rather than visual-quality dimensions of a single WIP
(Appendix~B). Internal scores and disagreement route candidate
discrepancies for human inspection and are not shown to participants as
aesthetic quality scores.

The model pipeline uses GPT-4o-mini, Gemini~2.0 Flash, and Claude 3.5 Sonnet. GPT-4o-mini
produces a concise visual observation; Gemini compares observations with specification
and reference context; Claude synthesizes the candidate analyses with the artwork and
reference hub. Reference pixels and annotations are passed with their metadata so output
can point to relevant regions. A complete analysis typically streams within 30--45
seconds. During consolidation, supervisor annotations and client input are combined with
model output, and the supervisor determines the final checklist.

The prototype also contains an important implementation limitation. Some evaluator
prompts used broad reviewer language such as ``what needs improvement'' and ``final
verdict,'' which could invite suggestions beyond an explicit project anchor. Appendix~F reproduces those prompts verbatim. The study therefore supports a human-authorized structured review
workflow, but it does not demonstrate that every model observation was evidence-bounded.
Section~\ref{sec:implementation-gap} returns to this limitation.

\section{SUMMATIVE STUDY}

We conducted one in-studio summative study with both senior- and junior-role
practitioners. The study examined three questions: whether MOONWALK supports cross-role
alignment (RQ1), produces executable review output (RQ2), and supports review awareness,
traceability, and agency (RQ3).

\begin{itemize}
    \item \textbf{RQ1 (Improve Collaboration):} Does MOONWALK improve intent alignment
    and shared standards between senior and junior artists?
    \item \textbf{RQ2 (Improve Outcome / Efficiency):} Does MOONWALK produce
    evidence-linked review output that junior artists can execute with less clarification?
    \item \textbf{RQ3 (Review Awareness, Traceability, and Agency):} Does MOONWALK help
    participants clarify criteria, identify artifact--intent gaps and blind spots, trace
    review decisions, and retain control over final judgments?
\end{itemize}

\subsection{Study Design}

\subsubsection{Participants}
We recruited 19 animation and VFX practitioners across two studio contexts (Table~1 in
Appendix~C). Senior-role participants ($n=10$) had 4--16 years of experience; junior-role
participants ($n=9$) had 0.5--3 years. Ten participants also took part in the formative
study. Participant roles and overlap are listed in Appendix~C.

\subsubsection{Tasks and Procedure}
Each session lasted approximately 40--60 minutes and took place in a studio setting.
After a video demonstration, researcher-guided walkthrough, and Q\&A, participants
experienced two within-subject conditions. \textbf{MOONWALK} provided the structured
workflow with reference retrieval, specification inspection, AI-assisted analysis, and
evidence-linked action output. \textbf{Chat-Only} exposed the conversational front end of
the same underlying AI without the Reference Hub, specification linkage, Artist Interpretation
panel, or checklist output. Both conditions used matched
production materials: four projects, each with one specification, four artworks, and four
references.

The same post-task questionnaire was completed by all 19 participants. Senior-role
participants were asked to attend especially to review criteria, evidence, and final
decision-making; junior-role participants were asked to attend especially to execution
clarity and clarification needs. Interviews then elicited role-specific perspectives on
the same workflow. Participants' \textbf{Existing Workflow} was collected after the task
as a retrospective ecological comparison and was not a third controlled condition. 

\subsubsection{Measures and Analysis}
\begin{figure}[t]
    \centering
    \includegraphics[width=0.8\linewidth]{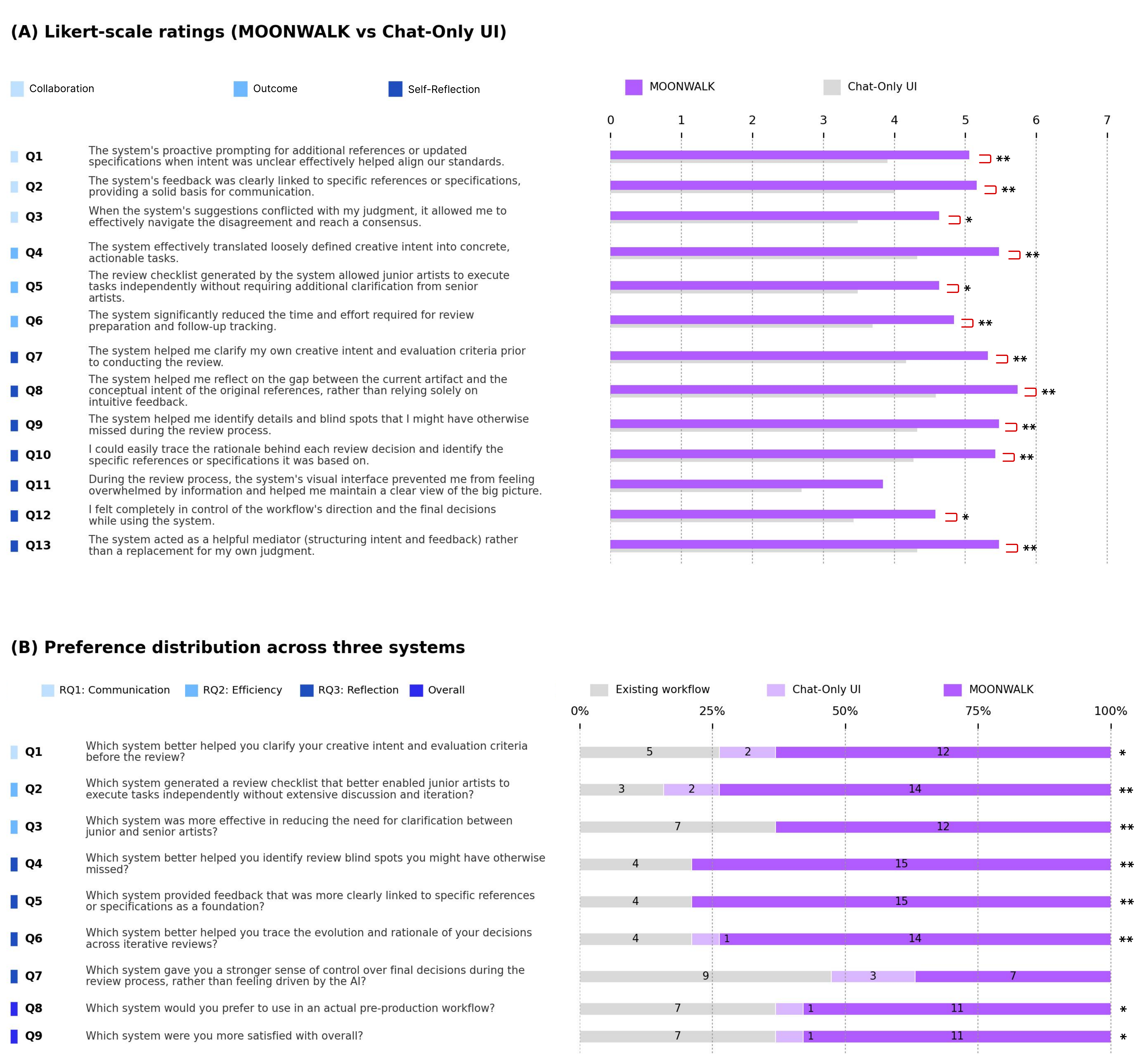}
    \caption{Survey results from the summative study. (A) Ratings for MOONWALK and Chat-Only on the 13 seven-point Likert items (reproduced in Appendix~E) across RQ1--RQ3, with significance assessed by one-sample Wilcoxon signed-rank test against the neutral midpoint (4). (B) Preference distributions across MOONWALK, Chat-Only UI, and Existing Workflows for 9 comparative questions (CQs), assessed by chi-square goodness-of-fit. *:~\textit{p}~<~.05 and **: \textit{p}~<~.01 report the paired MOONWALK vs.\ Chat-Only tests.}
    \label{fig:result}
\end{figure}

The 13-item questionnaire used a seven-point Likert scale (1 = Strongly Disagree,
7 = Strongly Agree) under the administered section labels \textit{Improve Collaboration}
(Q1--Q3), \textit{Improve Outcome / Efficiency} (Q4--Q6), and \textit{Improve
Self-Reflection} (Q7--Q13). A nine-item comparative questionnaire asked participants to
choose among MOONWALK, Chat-Only, and Existing Workflow. Appendix~E reproduces the
administered wording. Figure~\ref{fig:result} uses shortened item labels for readability;
those labels should not be read as alternate questionnaire wording.

For each Likert item we report (i) a one-sample Wilcoxon signed-rank test
of MOONWALK ratings against the neutral midpoint (4) and (ii) a paired
Wilcoxon signed-rank test of MOONWALK against Chat-Only ratings from the
same participants (zeros retained via the Pratt method; normal
approximation with continuity correction). For each three-way comparative question, we report a chi-square
goodness-of-fit test against a uniform preference distribution. This test indicates
whether the three-option distribution departs from uniformity; it is not a pairwise
significance test between MOONWALK and Existing Workflow. Interviews were analyzed by
three researchers using thematic analysis \cite{braun2006thematic}.

\section{RESULTS \& FINDINGS}

We report results for all 19 participants: one-sample Wilcoxon tests of
MOONWALK's Likert ratings against the neutral midpoint, chi-square
goodness-of-fit tests on three-way preferences, and interview themes
(Figure~\ref{fig:result}). MOONWALK was rated significantly above neutral
on 12 of 13 Likert items and received the plurality on eight of nine
comparative questions; sense of control (CQ7) was the exception.

\subsection{RQ1: Improving Collaboration \& Intent Alignment}

MOONWALK was rated significantly above neutral on all three RQ1 items: proactive requests
for additional references or specification updates when intent was unclear (Q1,
$p<.01$), linking feedback to specific references or specifications (Q2, $p<.01$), and
handling conflicts between system suggestions and personal judgment (Q3, $p<.05$). For
CQ1, 12 of 19 participants (63\%) chose MOONWALK for articulating creative intent and
review criteria, compared with 2 for Chat-Only and 5 for Existing Workflow ($p<.05$ for
the three-way goodness-of-fit test).

Participants valued keeping specifications, references, and review output in the same
record. P10 wanted ``folder documents, Zulip threads, ShotGrid feedback, and all oral
decisions'' available in one review context, while P2 noted that ``a brief note like `move it left' often encodes a
composition problem; the system surfaces and records that.'' P13 emphasized that evidence-linked output made it clearer what needed to be
fixed before escalating another question to a lead. Senior practitioners (P8, P9),
however, found some generated summaries ``too diplomatic'' and insufficiently incisive.

\subsection{RQ2: Improving Outcomes \& Efficiency}

MOONWALK was rated significantly above neutral on translating loosely defined intent into
actionable items (Q4, $p<.01$), producing checklists that junior artists could execute
without additional clarification (Q5, $p<.05$), and reducing participants' \emph{mental
effort in preparing for review and tracking follow-up actions} (Q6, $p<.01$). For CQ2,
14 of 19 participants (74\%) chose MOONWALK for junior-executable checklists, compared
with 2 for Chat-Only and 3 for Existing Workflow ($p<.01$). For CQ3, 12 of 19 (63\%)
chose MOONWALK for reducing senior--junior clarification, 0 chose Chat-Only, and 7 chose
Existing Workflow ($p<.01$).

Participants linked checklist executability to reference and specification linkage. They
estimated that juniors could independently resolve 70--80\% of foundational errors
(P3, P16, P18, P20), especially checks with an explicit target, such as scale proportion,
a specified color-temperature relation, or consistency with a supplied reference. P16
described the workflow as an interactive SOP for checking omissions before advancing to
dynamic animation. P13 also described using recorded rationale when a previously
approved direction was later questioned, while P12 noted that review history could expose
conflicts between current and earlier client instructions. These are participant
assessments, not logged longitudinal reductions in production time or rework. P5 further
estimated that roughly 99\% of revision cycles in their context originated from client
feedback, limiting any expected rework benefit to internal iteration.

\subsection{RQ3: Review Awareness, Traceability, \& Human Agency}

Within the questionnaire's administered \textit{Improve Self-Reflection} section,
MOONWALK was rated significantly above neutral on six of seven items (Q7--Q13; Q11
n.s.). Q8---``The system helped me reflect on gaps between the current artifact and the
original intent, rather than relying on intuitive judgment alone''---had the highest mean
across all 13 items ($M=5.74$, $p<.01$). We treat this as a reported outcome about
artifact--intent inspection, not as evidence that reflection is a causal system mechanism. Participants tied this to persistence and coverage: ``projects eventually
forget their own initial spec; the system never does'' (P4), and ``when
I'm fixated on color, the system flags that material and composition have
gone unchecked'' (P2).
For blind-spot identification (CQ4) and evidence-linked feedback (CQ5), 15 of 19
participants (79\%) chose MOONWALK and none chose Chat-Only in either question
(both $p<.01$).

Traceability (Q10) was significantly above neutral ($p<.01$), and 14 of 19 participants
(74\%) chose MOONWALK for tracking decision evolution (CQ6, $p<.01$). P13 explained
that when a review call was challenged, the reference that motivated it could already be
on record. On agency, Q13 (system as mediator) was above neutral ($p<.01$) and Q12
(sense of control) was above neutral ($p<.05$), but CQ7 was not significant: 47\% chose
Existing Workflow and 37\% chose MOONWALK ($p=.23$). P5 warned that habit erodes judgment: ``if you get used to it, you gradually surrender your own judgment---a system that is 90\% correct can make you careless about the remaining 10\%.'' P8 described artists without discriminating ability executing model output wholesale as ``a disaster if
followed across ten suggestions simultaneously.''

The Artist Interpretation panel let artists record their interpretation
before review, but completing it was not enforced as a gate. The visual
interface's ability to prevent information overload was the only Likert item that did not
reach significance (Q11, $p=.73$). Participants also requested visual post-revision
previews (P9, P15). Overall, 11 of 19 participants (58\%) preferred MOONWALK for a real
pre-production workflow (CQ8, $p<.05$), and 11 of 19 were more satisfied with it (CQ9,
$p<.05$); in both questions 7 selected Existing Workflow and 1 selected Chat-Only.

\section{DISCUSSION, LIMITATIONS, AND FUTURE WORK}

\subsection{Intent--Evidence--Action as a Coordination Problem}

Across the formative and summative studies, participants repeatedly valued the same
coordination properties: keeping the active specification and references available during
review, tracing a judgment to its rationale, and returning a prioritized action record to
the artist. Our evaluations indicate that pre-production review should be designed around
the continuity of intent, evidence, and action across a handoff. Similar coordination
problems may arise in game development \cite{begemann2024empirical} and other
role-differentiated creative production settings, but whether these patterns extend to adjacent domains like game development requires further empirical testing.

\subsection{Collaboration Through a Shared Review Record}

The collaboration problem is an information asymmetry between roles. Supervisors hold
project history, client context, and craft knowledge that may be compressed into a few
words of feedback; junior artists need enough of that context to act without repeatedly
interrupting the group lead (Section~6.1).

The prototype also stored the artist's interpretation with the submitted work, which
creates an opportunity for richer collaboration: a supervisor can see why a junior made a
choice before responding to the artifact. The study did not separately test an explicit
confirmation step in which the two roles negotiated that interpretation before review.
Future work should explicitly isolate whether adding a mutual confirmation step improves review outcomes.

\subsection{Different Roles Need Different Support Within the Same System}

A shared record does not imply identical assistance. Junior artists valued explicit
requirements, relevant references, bounded next actions, and completion conditions.
Senior practitioners already held much of the standard in mind and instead wanted direct
identification of missing evidence and the object or region requiring attention; several
found broad summaries ``too diplomatic.''

The same intent--evidence--action record can support these differences without splitting
the workflow into separate systems. For a \textbf{senior}, AI can request a missing
reference/specification, help identify the analysis target, retrieve prior decisions, and
organize annotations. For a \textbf{junior}, it can remind them of stated constraints,
compare the WIP with cited evidence, surface objective omissions, help locate the
applicable brief/spec, and capture why they intentionally deviated. That explanation then
travels with the WIP into supervisor review. While both roles access the same project record, supervisors require high-level discrepancy flags, whereas junior artists need specific, step-by-step revision guidance \cite{feedbackByDesign2026,aideation2025}.

\subsection{Agency and Mentorship in the Same Collaboration Loop}

Tailoring UI features to specific roles directly dictates how creative agency is maintained during automated assistance. The over-reliance concern in Section~6.3 echoes prior findings that
greater AI involvement can reduce cognitive engagement and diversity
\cite{chen2025assistance, doshi2024generative}. Human authorization helps keep
the final production decision with practitioners, but the quality of that decision still
depends on whether the model output is grounded in project evidence and whether the
artist can question it.

This concern connects directly to mentorship. Reducing routine clarification could free
senior time for higher-value teaching, but automation could also reduce the interactions
through which juniors learn tacit standards. A production deployment should therefore
separate routine evidence checking from mentorship: the former can be assisted, while the
latter requires continued senior--junior interaction and explanation.

\subsection{Visual-Native Review and Real-World Adoption}
\label{sec:discussion-adoption}

Participants found text-heavy output poorly matched to visual production and suggested a
post-revision preview image beside the checklist. Such a preview could make a proposed
change visually inspectable, but it was not implemented here. They also expected the
workflow to fit larger projects and the lead-to-junior boundary after creative direction
was established (P3, P4, P9, P18). The main adoption barrier was tool friction: a parallel
platform beside ShotGrid, Zulip, and existing pipeline tools would be difficult to sustain
(P9, P10, P19). Integration with production-management infrastructure is therefore a more
plausible deployment path than maintaining a separate review platform.

\subsection{Limitations and Future Work}
\label{sec:implementation-gap}

\textbf{Evaluation Scope and Methodological Trade-offs.}
Our evaluation was designed as a single-session, exploratory study to validate the immediate usability and feasibility of the MOONWALK workflow. While practitioner estimates (e.g., resolving 70--80\% of foundational issues) highlight strong subjective utility, they represent perceived efficacy rather than longitudinal production logs. Furthermore, to evaluate the holistic end-to-end interaction, the study prioritized comparing MOONWALK against a baseline chat interface rather than isolating individual systemic components (such as persistent structure vs. AI capabilities). Future controlled studies should incorporate non-AI structured baselines, counterbalance condition orders to eliminate sequence effects, and track long-term metrics (e.g., actual rework hours and clarification cycles) across multi-week production pipelines.

\textbf{Scope of the Evidential Record.}
MOONWALK’s analytical capabilities are structurally bounded by the completeness of its shared project record. Tacit decisions, unrecorded verbal exchanges, and external client messages remain invisible to the system unless explicitly imported. In longitudinal deployments, uncaptured context may compound across handoffs. Future work should prioritize low-friction context ingestion—such as review thread parsing, meeting transcript ingestion, and pipeline management tools (e.g., ShotGrid) webhooks—while rigorously maintaining asset provenance.

\textbf{From Prototype Instantiation to Production Deployment.}
MOONWALK operationalizes its framework through a specific set of interfaces, models, and review checkpoints. To generalize these insights, future work should evaluate how this intent--evidence--action loop scales across different studio cultures, production stages, and tool ecosystems. Promising technical extensions include diffusion-based visual revision previews and domain-specific perceptual models; however, such extensions must preserve the core boundary: AI should surface candidate observations, while actionable review decisions remain strictly traceable to shared evidence and authorized by human practitioners.
\section{CONCLUSION}
We presented an intent--evidence--action design framework for professional junior–supervisor artists workflow by MOONWALK, a working pre-production review system. Across a formative study and a within-subject in-studio evaluation, practitioners valued the integrated system's persistent specifications and references, evidence-linked review records, and executor-ready checklists; our evaluation revealed role-based differences in tool usage, limits on perceived control, and risks of over-reliance. These findings support MOONWALK as a structured alternative to unstructured conversational AI within the evaluated tasks and materials. They do not isolate the contribution of individual mechanisms, establish that AI is necessary beyond structured review support, or evaluate an explicit supervisor--artist interpretation checkpoint. More broadly, the work identifies a design opportunity for collaborative creative systems: preserve how articulated intent is grounded in evidence and translated into supervisor-approved actionable revision tasks, while keeping aesthetic authority and final judgment with professionals.

% ── References ───────────────────────────────────────────────────
\bibliographystyle{unsrt}
\bibliography{sample-base}

% ── Appendix ─────────────────────────────────────────────────────
\clearpage
\appendix
\section*{Appendix A: Technical Implementation Details}
\label{Appendix_Tech}

\paragraph{Project state.}
The evaluated implementation stores the structured brief as typed fields and each
reference with its category, priority tier, and annotation note in a per-project evidence
store. Analysis, review, and consolidation read the same state, so later runs use the
latest brief, reference notes, artist interpretation, and prior review decisions.

\paragraph{Eleven-dimensional analysis.}
A work-in-progress is analyzed across lighting, composition, color, style, perceptual
quality, sketch/line quality, specification faithfulness, controllability, consistency,
efficiency, and stability. Two heuristic checks are run for each dimension. They combine
image-derived signals with lexical alignment to the active brief and artist notes, then
use their internal confidence and agreement to organize candidate discrepancies for
human inspection. The resulting values are internal routing signals only.

\paragraph{Feature groups.}
The implementation uses five groups of image signals. Composition uses saliency-based
layout, horizon orientation, and center-bias cues; lighting and color use white-balance
and exposure cues; no-reference image quality uses sharpness and luminance-distribution
statistics; artifact checks detect blocking and banding; and style/prompt alignment uses
CLIP ViT-B/32 similarity. These signals are combined with the active specification,
reference notes, and the artist's submitted interpretation. Different checks use
complementary vocabularies and contradiction detection so that disagreement can identify
ambiguous cases for supervisor attention.

\paragraph{Three-model synthesis and output.}
Each candidate dimension passes through the evaluated three-model pipeline. GPT-4o-mini
produces a concise visual observation; Gemini 2.0 Flash compares it with the active
specification and reference context; and Claude 3.5 Sonnet synthesizes the supported analyses
with the artwork and reference hub. Reference pixels and metadata are passed directly so
the output can cite a visual region rather than only a filename. The full analysis
typically arrives within 30--45 seconds. Final consolidation combines supervisor
feedback, client input, and supported observations into a prioritized checklist with
completion criteria. Supervisor input is authoritative in conflicts, and every issued
item must be grounded in the brief, a specific reference, a prior decision, or explicit
human judgment. Canvas annotations are mapped back to source resolution and versioned.

\section*{Appendix B: Analysis Dimensions Definitions}
\label{app:analysis_dimensions}

The prototype's eleven analysis dimensions fall into two groups. Seven are
artifact- and evidence-facing: \emph{lighting}, \emph{composition},
\emph{color}, and \emph{style} (image signals checked against annotated
references and the brief); \emph{perceptual quality} and \emph{sketch/line
quality} (no-reference sharpness, luminance, and artifact statistics); and
\emph{specification faithfulness} (lexical and CLIP alignment to the
active brief). The other four are inherited from a generation-evaluation
rubric used earlier in development. Two of them are defined in the
implementation as response-level measures: \emph{efficiency} tracks
response length and \emph{stability} tracks agreement across repeated
responses. \emph{Controllability} and \emph{consistency} are retained from
the same rubric and are likewise not visual properties of a single WIP
frame. All four inform only internal routing, are never surfaced to
participants, and carry no construct-level claim in this paper; we report
them for completeness of the released implementation.

\section*{Appendix C: Formative Study — Participants \& Procedure}
\label{app:formative_procedure}

We recruited 12 practitioners (1--16 YoE, $M=4.96$) across two animation and VFX
studios through personal referrals: directors ($n=2$), a supervisor ($n=1$), artists
($n=8$; three senior-role and five junior-role artists), and a PM ($n=1$). The formative
participants were P2, P3, P6, P8, P9, P11, P12, P13, P15, P16, P17, and P18 in
Table~\ref{Demography}. Production contexts spanned commercial advertising, character
animation, virtual production, and film/TV VFX. Sessions lasted 30--60 minutes; two
participants completed additional 30--45-minute follow-up sessions focused on error
typology, reference alignment failure, and junior onboarding. Interviews were
audio-recorded, transcribed, and analyzed using thematic analysis~\cite{braun2006thematic};
an author with prior studio experience developed the initial coding framework, and themes
were iteratively refined across four co-authors. The sample included two animation
directors (P6, P8), one VFX supervisor (P9), a technical artist lead (P2), a CG lead
(P3), a senior concept artist (P11), five junior artists (P12, P13, P16, P17, P18), and a
project manager (P15). Full interview questions are provided in Appendix~D.

\begin{table*}[t]
\centering
\caption{Demographic Details of Participants}
\label{Demography}
\resizebox{\textwidth}{!}{
\begin{tabular}{|c|c|l|c|c|c|}
\hline
\textbf{ID} & \textbf{Years of Exp.} & \textbf{Job Title} & \textbf{Role Level} & \textbf{Formative (12)} & \textbf{Summative (19)} \\
\hline
P1 & 7.5 & Pre-Production Lead (CG Assets) & Senior &  & \checkmark \\
P2 & 8 & Technical Artist Lead (Environment) & Senior & \checkmark & \checkmark \\
P3 & 6 & CG Lead (Ads+Products) & Senior & \checkmark & \checkmark \\
P4 & 5 & CG Lead (Animation) & Senior &  & \checkmark \\
P5 & 12 & Animation Director (Executive) & Senior & & \checkmark \\
P6 & 10 & Animation Director (Product) & Senior & \checkmark & \checkmark \\
P7 & 7 & Unreal Art Director (Ads) & Senior &  & \checkmark \\
P8 & 4 & Animation Director (Ads+Game) & Senior & \checkmark & \checkmark \\
P9 & 16 & VFX Supervisor (Films+TV shows) & Senior & \checkmark & \checkmark \\
P10 & 11 & CG Supervisor (Character) & Senior &  & \checkmark \\
P11 & 5.5 & Concept Artist (Design) & Senior & \checkmark & \\
P12 & 3 & Concept Technical Artist (Character+Scene) & Junior & \checkmark & \checkmark \\
P13 & 2 & AI Concept Artist (Character+Scene) & Junior & \checkmark & \checkmark \\
P14 & 0.5 & Concept Intern (Character+Scene) & Junior &  & \checkmark \\
P15 & 2 & Project Manager (Character+Motion) & Junior & \checkmark & \checkmark \\
P16 & 1 & Style Frame Artist (Concept+3D) & Junior & \checkmark & \checkmark \\
P17 & 1 & Style Frame Artist (Scene+AI) & Junior & \checkmark & \\
P18 & 1 & Style Frame Artist (Motion+3D) & Junior & \checkmark & \checkmark \\
P19 & 2 & Lighting \& Composition Artist (Character) & Junior &  & \checkmark \\
P20 & 1 & Animator (Ads+Music Video) & Junior &  & \checkmark \\
P21 & 1 & 3D Modeling Artist (Product+Game) & Junior &  & \checkmark \\
\hline
\end{tabular}%
}
\end{table*}

\FloatBarrier

\section*{Appendix D: Formative Study — Interview Questions}
\label{app:formative}

\noindent\textit{Goal: Understand how directors, supervisors, and artists communicate intent and where the review cycle breaks down.}

\subsection*{Part 1 | Participant Background (\textasciitilde2 min)}
\begin{enumerate}[label=\textbf{A\arabic*.}]
  \item What is your primary role and responsibilities in your current project? How much time do you spend reviewing and giving feedback?
  \item What types of productions have you worked on? (2D/3D, TV/film/commercial)
\end{enumerate}

\subsection*{Part 2 | Current Review \& Communication Workflow (\textasciitilde5 min)}
\begin{enumerate}[label=\textbf{B\arabic*.}]
  \item Walk me through what happens from the moment an artist sends you a version to the moment you return comments. What tools do you use? What format does feedback take?
  \item How many rounds does a shot typically go through? Which part takes the most time---watching, articulating the issue, or re-working after a misunderstanding?
  \item When you review a frame, do you have a mental checklist (composition, lighting, color, style consistency, continuity)? Have these criteria ever been written down?
  \item How do you explain \textit{why} something needs to change---abstract language or specific parameters?
\end{enumerate}

\subsection*{Part 3 | Pain Points (\textasciitilde4 min)}
\begin{enumerate}[label=\textbf{C\arabic*.}]
  \item What kinds of misunderstandings come up most often---different interpretations of tone, different standards for ``done,'' or different readings of the same instruction?
  \item Walk me through a specific example where feedback went back and forth many times. Where did it get stuck?
  \item Which part of your current review workflow is most draining?
\end{enumerate}

\subsection*{Part 4 | Expectations for an Assistive System (\textasciitilde3 min)}
\begin{enumerate}[label=\textbf{D\arabic*.}]
  \item If a system could run a preliminary check before you review, what would you want it to look for?
  \item At which moment in the workflow would you want the system to step in?
  \item How would you judge whether this system is actually helpful?
  \item Do you have concerns about AI assistance (creative constraint, incorrect evaluation, privacy)?
\end{enumerate}

\section*{Appendix E: Summative Study - Details and Questions}

\subsection*{1. Questionnaire Questions for Summative Study}
\label{app:questionnaire}

\paragraph{Reporting note.}

\textit{After-Task Questionnaire} (13 items). 1--7 Likert Scale: Strongly Disagree $\rightarrow$ Strongly Agree.

\paragraph{Section 1 | Improve Collaboration}
\begin{enumerate}
  \item The system's proactive requests for additional references or spec updates when intent was unclear effectively helped align our standards.
  \item The system's feedback was clearly linked to specific references or specifications as a basis for communication.
  \item When the system's suggestions conflicted with my judgment, the conflict-resolution loop allowed me to handle disagreements effectively.
\end{enumerate}

\paragraph{Section 2 | Improve Outcome / Efficiency}
\begin{enumerate}[resume]
  \item The system effectively translated loosely defined creative intent into concrete, actionable review items.
  \item The review checklist generated by the system was actionable enough for junior artists without requiring additional clarification.
  \item The system reduced my mental effort in preparing for review and tracking follow-up actions.
\end{enumerate}

\paragraph{Section 3 | Improve Self-Reflection}
\begin{enumerate}[resume]
  \item The system helped me clarify my own creative intent and review criteria before conducting feedback.
  \item The system helped me reflect on gaps between the current artifact and the original intent, rather than relying on intuitive judgment alone.
  \item The system helped me identify details and blind spots I might have otherwise missed.
  \item I could easily trace the rationale behind each review decision and its supporting references or specs.
  \item During review, the system's visual interface prevented me from feeling overwhelmed by information.
  \item I felt fully in control of the workflow direction and final decisions.
  \item The system acted as a mediator (helping to structure intent and feedback), rather than replacing my judgment.
\end{enumerate}

\bigskip
\textit{Comparative Questionnaire} (9 items). A/B/C preference comparison.
\begin{enumerate}
  \item \textbf{[RQ1]} Which system better helped you clearly articulate your creative intent and review criteria?
  \item \textbf{[RQ2]} Which system's generated review checklist allowed junior artists to execute independently?
  \item \textbf{[RQ2]} Which system more effectively reduced the need for clarification between junior and senior artists?
  \item \textbf{[RQ3]} Which system better helped you identify review blind spots?
  \item \textbf{[RQ3]} Which system's feedback was more clearly linked to specific references or specifications?
  \item \textbf{[RQ3]} Which system better helped you track the evolution and rationale of decisions across iterative reviews?
  \item \textbf{[RQ3]} Which system gave you a greater sense of control over final decisions?
  \item \textbf{[Overall]} If used in a real pre-production workflow, which system would you prefer?
  \item \textbf{[Overall]} Which system did you find more satisfying overall?
\end{enumerate}

\subsection*{2. Interview Questions for Summative Study}
\label{app:interview}

\paragraph{Part 0 | Overall Experience}
\begin{enumerate}
  \item Share your overall experience using MOONWALK from start to finish. Compared to your previous workflow, what do you consider the most fundamental difference?
\end{enumerate}

\paragraph{Part 1 | RQ1 --- Improve Collaboration}
\begin{enumerate}
  \item \textbf{Concretizing Intent.} Was the system effective in helping you clarify and concretize your creative intent? How was structuring it through the system different from relying on memory or chat logs?
  \item \textbf{Communication Standards.} Did the system substantially improve communication efficiency between senior and junior artists?
  \item \textbf{Evidence-Based Dialogue.} The system links each review item to a specific reference or spec. How did this design help you communicate with other artists?
\end{enumerate}

\paragraph{Part 2 | RQ2 --- Improve Outcome / Efficiency}
\begin{enumerate}
  \item \textbf{Reducing Clarification Cost.} Could a junior artist execute the generated checklist directly without asking follow-up questions?
  \item \textbf{Reducing Rework Risk.} Did the system help reduce the risk of late-stage decision changes? Can you give a concrete example?
  \item \textbf{Tracking Decisions.} Did the ability to track decision evolution across iterative reviews help address inconsistencies in review standards?
\end{enumerate}

\paragraph{Part 3 | RQ3 --- Improve Self-Reflection}
\begin{enumerate}
  \item \textbf{Guided Reflection.} Did MOONWALK prompt structured reflection on project requirements and areas for improvement in current artifacts?
  \item \textbf{Discovering Blind Spots.} Did the system's feedback help you identify issues you might have missed? Did the system's judgment align with your own?
  \item \textbf{Conflict \& Sense of Control.} In situations of user--AI disagreement, how did you feel about your sense of control over the review process?
  \item \textbf{Mediator Role.} Do you feel the system successfully fulfilled the role of a coordinator/mediator?
\end{enumerate}

\paragraph{Part 4 | Wrap-up}
\begin{enumerate}
  \item \textbf{Challenges.} Were there any challenges or frustrations? Were there features notably missing?
  \item \textbf{Future Integration.} How would you integrate MOONWALK into your existing workflow? What type of studio or team size would be best suited?
\end{enumerate}

\section*{Appendix F: Details of Multi-Agent Prompts}

\subsection*{1. Role-Based Evaluator Prompts}
\label{app:role_prompts}

\paragraph{Implementation boundary.}
These prompts are reproduced verbatim from the evaluated
prototype; as Section~7.6 discusses, some use broader reviewer language
than the framework's evidence-bounded AI role. In
particular, Agent~A can ask for a ``critical issue'' and its ``negative impact,'' and
Agent~C can request target values and a ``verdict'' without an explicit instruction to
abstain when project evidence is missing. We therefore treat this as an implementation
limitation, not as evidence that all model observations were grounded. Supervisor
review remained the final authorization step; Section~\ref{sec:implementation-gap} discusses the consequence for our claims.

Each evaluator agent is initialized with a role-specific system prompt. The \textbf{Base} prompt provides shared evaluation principles; role prompts extend it with persona-specific focus areas.

\paragraph{Base Evaluator}
\begin{Verbatim}[
  fontsize=\footnotesize,
  breaklines=true,
  breakanywhere=true,
  frame=single,
  framesep=6pt
]
You are evaluating an image compared to a user's reference image.
Your goal is to provide honest, specific, and actionable feedback.
Focus on what is working well and what needs improvement.
Always respond in the same language as the user's request.
Be concise but thorough. Avoid vague praise or generic criticism.
\end{Verbatim}

\paragraph{Three-Agent Debate Prompts}

\noindent Agent~A (OpenAI) provides visual observation; Agent~B (Gemini) provides a technical-versus-spec comparison; Agent~C (Claude) synthesizes both into actionable directives.

\paragraph{Agent A --- Visual Observer (System Prompt)}
\begin{Verbatim}[
  fontsize=\footnotesize,
  breaklines=true,
  breakanywhere=true,
  frame=single,
  framesep=6pt
]
You are a senior VFX visual reviewer evaluating "{metric_name}".
Task: identify the most critical issue---where it appears in the frame
and what negative impact it has on the overall visual experience.
Requirements: include specific quantitative values (angle, ratio,
color value, contrast). Under 130 characters. No markdown.
\end{Verbatim}

\paragraph{Agent B --- Technical Reviewer (System Prompt)}
\begin{Verbatim}[   fontsize=\footnotesize,   breaklines=true,   breakanywhere=true,   frame=single,   framesep=6pt ]
You are a senior VFX technical reviewer evaluating "{metric_name}"
against the Director's Spec and References. Identify the specific gap
in the format "Reference has X as ___, current work is ___, gap is
approximately ___". Provide 1-2 immediately actionable steps.
Under 130 characters. No markdown.
\end{Verbatim}

\paragraph{Agent C --- Supervisor / Final Verdict}
\begin{Verbatim}[   fontsize=\footnotesize,   breaklines=true,   breakanywhere=true,   frame=single,   framesep=6pt ]
You are a VFX Senior Supervisor. Synthesize the two agents'
observations and produce 2-3 specific improvement directives.
Each directive must use the format: "Adjust [specific parameter]
from [current value] to [target value]". If agents disagree,
state the disagreement in one sentence, then give the verdict.
Under 200 characters. No markdown.
\end{Verbatim}

\subsection*{2. Spec Summary Prompt}
\label{app:spec_summary}

After all per-dimension debates complete, Claude synthesizes a holistic summary of how well the artist's references align with the director's spec. Runtime variables (\texttt{\{brief\}}, \texttt{\{hub\_refs\}}, \texttt{\{artist\_refs\}}, \texttt{\{reflection\}}, metric aggregates) are injected before the call. System prompt: \textit{``You are a VFX Supervisor providing a Spec + Reference alignment summary of the artist's work. Under 300 characters. No markdown.''}

\end{document}